\documentclass[12pt]{article}

\usepackage{tikz}

\newcommand{\be}{\begin{equation}}
\newcommand{\ee}{\end{equation}}
\newcommand{\ba}{\begin{eqnarray}}
\newcommand{\ea}{\end{eqnarray}}

\newcommand{\V}{\varphi}
\newcommand{\Tr}{\rm Tr}

\newcommand{\G}{\Gamma}

\newcommand{\e}{\epsilon}

\begin{document}
\hoffset=-.4truein\voffset=-0.5truein
\setlength{\textheight}{8.5 in}
\begin{titlepage}

\begin{center}
\hfill { 
January 2023}\
\vskip 16 mm

{\large \bf Should we worry about renormalons  in the $\epsilon$-expansion? }
\vskip .6 in

 {\bf E. Br\'ezin$^{a)}$}
\end{center}
\vskip 5mm
\begin{center}

{$^{a)}$ Laboratoire de Physique de l'Ecole normale sup\'erieure, ENS,
Universit\'e PSL, CNRS, Sorbonne Universit\'e, Universit\'e de Paris, F-75005 Paris, e-mail: brezin@lpt.ens.fr} \\
\vskip 5mm
{\bf{ Submitted for publication in a book commemorating\\ Michael Fisher \\
({\it{Amnon Aharony, Ora Entin, David Huse, Leo Radzihovsky, editors}})}}

 \end{center}

 \vskip 2cm
{\bf Abstract} 
Turning the divergent $\e$-expansion into a  numerically sensible algorithm, relies on the knowledge of the behaviour of the large order contributions. Two different pictures are known to compete  there. The first one was based on Lipatov's instantons, which is known to deal with  the multiplicity of Feynman diagrams which grows factorially at high orders. However this was challenged by 't Hooft's renormalons who pointed out that renormalization could yield a similar growth through one single diagram. We study here a well-known model, the $O(N)$ model, in the large $N$ limit.  The reason for returning to this familiar model, is that it deals with diagrams known to give renormalon effects.Through an explicit   analytic result, we find no sign of a non-analyticity of perturbation theory due to these renormalons.           
   
  \end{titlepage}

  \vskip 3mm

 \section{Introduction}

A little more than fifty years ago  K.Wilson and M.Fisher introduced the  $\e$-expansion in the celebrated article \cite {WF} {\it{ Critical Exponents in 3.99 Dimensions}}.  This article has had a considerable influence both in the area of critical phenomena and in quantum field theory in general. However in practice,  if the first two terms of the expansion provided often a reasonable approximation to the measured indices in dimension three, the situation deteriorated if one  pushed the procedure to higher orders \cite{BZN}. Clearly the expansion looked divergent, it is believed to be at best asymptotic, and in fact  limited in its applicability.  Of course summation procedure, such as Borel transforms, Pad\'e approximants,  could be tried, but in the absence of further indication,  it was a blind shot. Therefore at the time analytic calculations, and computer simulations, appeared to be limited in their   ability of reproducing precision measurements by the renormalization group approach.

The situation changed significantly in 1977 with Lipatov approach  to the characterization of large orders in pertubation theory \cite{Lipatov}. His instanton method , which he developed for the field theory $g\phi^4$ in dimension four, predicted a large order behaviour of the perturbation expansion of the various correlation functions, $\beta$-function, etc, of the form $g^k k! (-a)^k k^b c$  for large $k$,  with calculable coefficients $a,b,c$ .   I shall recall below why this information, the explicit knowledge of $a,b,c$, is essential to extract a numerically sensible result from such a divergent  series.  With Le Guillou and Zinn-Justin \cite {BLZ} we first checked these results in lower dimensions. For instance the one-dimensional quartic anharmonic oscillator had been extensively studied by Bender and Wu \cite {BW} who had  found, on the basis of a WKB method, that indeed perturbation theory diverged with a $k!$ growth as above. We verified that the instanton method reproduced exactly  what they had found.  Extension to field theory in higher dimensions was similar \cite{BLZ} except that the instanton solution is not  known in analytic  form except in dimension four \cite{BP}. In dimensions lower than four the interpretation of these results is  simple : at a given order all Feynman diagrams have the same sign (we are considering an Euclidean field theory)  and their number grows proportionaly to $k!$. The situation is less transparent in dimension four since  diagrams require counterterms   and subtractions. This is the central point of the investigation that we try to carry in this paper. 

The knowledge  of the large order behaviour of the perturbative expansion was the source  of considerable improvements in the ability to get a sensible answer from those rapidly divergent series \cite {LZ}.  In practice it led to numerically convergent algorithms  developed around Zinn-Justin and collaborators\cite{LZ, GZ}, Borel transform of the series,  followed by conformal mappings relying on the explicit knowledge of the coefficients $a.b,c$ here above, were among the techniques that they used and examples, such as the anharmonic oscillator, revealed that the divergent perturbation series could lead, after those mappings, to many digits exact results.   

In our 1977 paper \cite {BLZ} we tried  to carry   the large order  knowledge of perturbative series to the $\e$-expansion. In most procedures one has to perform a double expansion in $g$ and $\e$. Since the fixed point $g^{*}$ is of order $\e$ one has to consider at order $k$, terms of order $g^k, g^{k-1}\e, g^{k-2} \e^2, \cdots$ . Fortunately the well-known technique of {\it{minimal subtraction}} allows one to avoid this double-expansion. In this renormalization scheme \cite{'t W} by definition  the counter-terms  contain only poles in $\e$ without any finite part.  As a result one obtains the renormalization group $\beta$ and $\gamma$ functions in dimension $d$ from their four-dimensional counterpart
\ba  &&\beta_d(g) =-\e g + \beta_4 (g) \nonumber\\
&&\gamma_d(g) = \gamma_4 (g) 
\ea
 The fixed point $g^{*}$ is then $\e$-expanded from 
 \be \e g^{*} = \beta_4(g^{*})\ee
($\beta_4(g)$ is a series starting at order two),  and the critical exponent $\eta$, which is (twice) the conformal anomaly of the $\phi$-field is then expanded  in $\e$  from
 \be \eta =  \gamma_4(g^{*})\ee
Therefore computing the $\e$-expansion is  reduced to  four-dimensional  calculations. Using the 4D large-order calculation from Lipatov method we obtained an estimate of the asymptotic orders in the $\e$-expansion \cite{BLZ} which was again of the $k!$-type as before.  Then  Zinn-Justin and co-workers based a summation procedure on this large order behaviour \cite{LZ, GZ}, and the process looked nicely convergent : at least adding one more order in the $\e$-expansion improved the previous result, instead of destroying it as the  straight expansion does. 

The confidence in this process  was severely affected after 't Hooft's discovery of a phenomenon \cite{'t}, now called {\it{renormalons},} which appears only in renormalizable theories such as $\phi^4$ in dimension four (and not in lower dimensions). His argument was based on the fact that one single diagram at order $k$ could be proportional to $k!$ , whereas the $k!$ , for $d<4$,  resulted from their multiplicity. In a renormalizable theory a diagram, such as the (renormalized) bubble diagram, grows logarithmically for a large external momentum. A  repeated insertion of such diagrams leads to an integral with a log at k-th power, which gives after integration a $k!$.  Clearly this shed doubt on Lipatov's estimate for large order, although 't Hooft's argument didn't show explicitely that it was wrong. It was not really clear  either how those renormalons affected
quantitatively the actual calculation. 

I have remained puzzled by this problem since then, wondering whether the ambitious RG  work based on the $\e$-expansion, was simply an approximation which should not be pushed too far.  I want to present here a simple well-known problem, namely the $O(N)$-model in the large-N limit. The reason for returning to this familiar model is that it is clearly a candidate for showing up renormalons effect, whereas this time we do not need an instanton asymptotics since there are only a few diagrams of a given order. I do not pretend that it solves the instanton-renormalon competition, but I wonder on the basis of this calculation whereas  renormalons affect the perturbation series as badly as one could think. 

\section  {The $O(N)$ model in dimension four}

This, half a century-old, model consists of an $N$-component order parameter $\phi_a, a=1,\cdots, N$,  with an interaction invariant under $O(N)$ \cite {MM}. This model has been studied by hundreds of authors and elaborate techniques have allowed to compute several orders in a $1/N$ expansion. We will limit ourselves here to the leading large $N$ terms   for the RG functions $\beta(g)$ and $\gamma(g)$ in the minimal subtraction scheme which allows to make easy contact with the $\e$-expansion.  The reason for our interest is that $\gamma(g)$ is of order  $1/N$ and the expansion in powers of $g$ of this leading term involves exactly the diagrams which have been identified as generating renormalon singularities in the perturbative expansion.  With the help of previous results on critical indices we will try to understand what the renormalons do to this expansion.

The model is given by an (Euclidean) action 
\be S = \int d^4 x [ \frac{1}{2} \nabla \phi_a \nabla \phi_a + \frac{1}{2} m_o^2 \phi^2 + \frac{g_0}{4! N  }( \phi^2)^2 ] \ee
which one could regularize by an ultra-violet cut-off. For the  reasons  mentioned above we prefer here  the dimensional regularization  by going to $4-\e$ dimensions, and then renormalize minimally by a coupling constant renormalization $Z_1$ and a field rescaling $\phi =\sqrt Z \varphi$. We work in the massless theory (critical temperature) and the action in terms of the  renormalized field reads (omitting mass counterterms)
\be S = \int d^d x [ \frac{Z}{2} \nabla \varphi_a \nabla \varphi_a  + \mu^{\e}\frac{g Z_1}{4! N  }( \varphi^2)^2 ] \ee
Varying $\mu$ at fixed bare theory we obtain the standard renormalization group functions of the Callan-Symanzik equation \cite{CZ}
\be  \e g + \beta(g)(1 + g \frac{d}{dg} \ln (Z_1/Z^2) )=0 \ee
\be \gamma(g) = \beta(g) \frac{d}{dg} \ln (Z) \ee
We are interested in computing the leading terms in a $1/N$ expansion of $Z_1$ which is $O(N^0)$ and $Z $  since $Z= 1+ O(1/N)$. 

\subsection {The $\beta$-function at order $N^0$}
The coupling constant involving an explicit factor $1/N$, the leading terms maximize diagrams with internal index-loops which provide a compensating factor $N$. For the two-point function the leading diagrams are of order $1/N$ and the field renormalization $Z = 1 + O(1/N)$. For the four-point function the leading diagrams consist of a string of "bubbles" all of order $1/N$ and this yields a vertex renormalization $Z_1$ of order $N^0$ which we now compute. 

Let us begin with a bubble with external momentum $p$
\be \label{bubble} B(p) = \int \frac{d^d q}{q^2 (p-q)^2} \ee
where we use the convention of omitting the usual geometric factor $ \frac{2\pi^{d/2}} {(2\pi)^d\Gamma(d/2) } $ included into a rescaling of $g$  that will be implied ; we take the scale factor $\mu$ as unit of momentum.
 \setlength\unitlength{1mm}

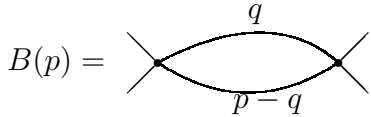
\begin{picture}(100,20)(-2,-2)
\put(20,0){\circle*{1}}
\put(16,4){\line (1,-1){4}}
\put(16,-4){\line (1,1){4}}
\put(32,6){$q$}
\put(30,-6){$p-q$}

\qbezier(20,0)(35, 8)(44,0)
\qbezier(20,0)(32, -8)(44,0)
\put(44,0){\line (1,1){4}}
\put(44,0){\line (1,-1){4}}

\put(44,0){\circle*{1}}

\put(0,-1) {$ B(p) =$}
\end {picture} 
\vskip 1cm

A standard calculation (using a Feynman parameter)  yields
\ba \label{B}&&B(p) = a(\e) \frac{p^{-\e}}{\e} ,\nonumber \\ &&a(\e) = \frac{1-\e/2}{1-\e}\  \frac {\G^3(1-\e/2) \G(1+ \e/2)}{\G(1-\e)} = 1+\frac{\e}{2} + O(\e^2) \ea
i.e.
\be B(p) = \frac{1}{\e} + 1/2 - \ln{p }\ee

Then the four-point function, for external indices a,a,b,b is given by a geometric series

 \setlength\unitlength{1mm}
\begin{picture}(100,20)(-2,-2)
\put(10,0){$..$}
\put(13,0){\line(1,-1){4}}
\put(6,4){\line (1,-1){4}}
\put(6,-4){\line (1,1){4}}
\put(13,0){\line(1,1){4}}
\put(9,-1) {$\cdots$}
\qbezier(27,0)(32, 6)(39,0)
\qbezier(27,0)(32, -6)(39,0)
\put(39,0){$...$}

\put(23,0) {$...$}

\put(42,0){\line(1,-1){4}}
\put(20,4){\line (1,-1){4}}
\put(20,-4){\line (1,1){4}}
\put(42,0){\line(1,1){4}}

\put(17,-1){$+$}

\put (46,-1){$+$}
\qbezier(57,0)(64, 8)(69,0)
\qbezier(57,0)(64, -8)(69,0)
\put(69,0){$...$}
\put (54,0){$...$}

\qbezier(72,0)(77, 8)(84,0)
\qbezier(72,0)(77, -8)(84,0)
\put (84,0){$...$}

\put(50,4){\line (1,-1){4}}
\put(50,-4){\line (1,1){4}}
\put(91,4){\line (-1,-1){4}}
\put(87,0){\line (1,-1){4}}
\put (90,-1) {$ + \cdots$}
\put(-10,-1) {$\G^{(4)} =$}
\end{picture}
\vskip 5mm

\ba  &&-\frac{1}{2}\G^{(4)} = \frac{1}{N 6} gZ_1-\frac{g^2 Z_1^2}{N 6^2} B(p) + \frac{g^3 Z_1^3 }{N 6^3} B^2(p) +\cdots \nonumber\\
&& = \frac{1}{N}\frac{gZ_1/6}{ 1+ gZ_1/6 \ B(p) } = \frac{1}{N}\frac{g/6}{ 1/Z_1+ g/6 \ B(p) } +O(1/N^2) \ea
Note that the Z factor omitted here would give only a  $1/N^2$ contribution. 
Then taking 
\be 1/Z_1= 1- \frac {g}{6\e} \ee
which satisfies the minimal subtraction rule, we obtain a finite $\G^{(4)}$ in the limit $\e \to 0$,   at order zero in $1/N$ . The $\beta$-function follows immediately
\be 0 = \e g + \beta(g) [ 1 + \frac{g}{6\e -g} ]\nonumber \ee
i.e.
\be \label{beta} \beta(g) = - \e g + \frac{1}{6} g^2 + O(1/N) \ee
One can check this result which is valid to all orders in $g$ , but zeroth order in $1/N$, with the literature. We copy from Zinn-Justin's book \cite{ZJ} in which he used the minimal subtraction scheme
\be \beta = -\e g + \frac{N+8}{6N}g^2-\frac {3N+14}{12 N^2}g^3 +\cdots \ee
and higher terms in $g$ are of order $1/N$, which agrees with (\ref{beta}) when $N$ goes to infinity. 
\subsection {The two-point function}
We are now considering the diagrams for the inverse two-point function $\G^{(2)}(p)$. 

 \setlength\unitlength{1mm}
\begin{picture}(100,20)(-2,-2)

\put(29,-2) {$p$}
\put(28,0){\line (1,0){5}}

\put(33,0) {$...$}

\put(39,11){$p-q$}
\qbezier(33,0)(47, 17)(65,0)

\qbezier(37,0)(50, 8)(60,0)
\qbezier(37,0)(50, -8)(60,0)
\put(65,0){\line (1,0){4}}
\put(60,0){$...$}
\put(-3,-1) {$ \G^{(2)}(p) = Zp^2 -\ $}

\end{picture}
\vskip 1cm

 At order $g^2$ we have one inserted bubble diagram, thus 
\be \G^{(2)} (p) = Z p^2 - 2\frac{(gZ_1)^2}{ N 6^2} \int d^d q \frac {B(q)}{(p-q)^2}  + O(g^3, 1/N^2) \ee
The integral over $q$ diverges as $-\frac{1}{8\e}$, we will compute explicitly the integral with a string of bubbles of arbitrary length  herefafter.  So at order $g^2$  
\be Z=1-\frac{1}{144 N } \frac{g^2}{\e} + O(g^3, 1/N^2) \ee
giving
\be \gamma(g) = \frac{g^2}{72N} +O(g^3, 1/N^2) \ee.

Tot all orders in $g$  the string of bubbles gives 
\be \label{S}  \G^{(2)} (p) = Z p^2 -\frac{2}{N} \sum _{k=1} (\frac {-gZ_1}{6})^{k+1} \int d^d q \frac {B^k(q)}{(p-q)^2} +O(1/N^2) \ee
We will sum the series later, but it is interesting to study the finite order $k$ dealing thus with the integral
\be \label{I} I_k =  \int d^d q \frac {B^k(q)}{(p-q)^2}  = \frac{a^k(\e)}{\e^k } \int d^dq \frac {q^{-k\e}}{(p-q)^2}   \ee
with $B(q)$ and $a(\e)$ given in  ({\ref{B}}).

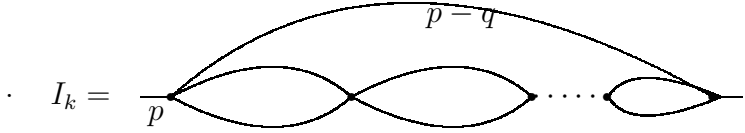
 \setlength\unitlength{1mm}
\begin{picture}(100,20)(-2,-2)
\put(22,0){\circle*{1}}
\put(46,0) {\circle*{1}}
\put(70,0) {\circle*{1}}
\put(80,0) {\circle*{1}}
\put(94,0) {\circle*{1}}
\put(6,-1) {$I_k =$}
\put(18,0){\line(1,0){4}}
\put(94,0){\line(1,0){4}}
\put(19,-3) {$p$}
\put(56,10){$p-q$}
\qbezier(22,0)(50, 25)(94,0)
\qbezier(22,0)(37, 8)(46,0)
\qbezier(22,0)(37, -8)(46,0)
\qbezier(46,0)(61, 8)(70,0)
\qbezier(46,0)(61, -8)(70,0)
\put(70,-1) {$\cdots \cdots$}. 
\qbezier(79,0)(82, 5)(94,0)
\qbezier(79,0)(82, -5)(94,0)

\end{picture}

\vskip 1cm
It is interesting to compute $I_k$ explicitly to understand what happens at higher $k$'s. Standard techniques give
\ba \label{Integral} I_k =&& -\frac{k}{4(k+1)} \frac{a^k p^{2-\e(k+1)}}{\e^k} \frac{\G^2(1-\e/2) \G(1-\e(k+1)/2)\G(1+\e(k+2)/2)}{ \G(1-\e(k+2)/2) \G(1+\e k/2)}\nonumber \\&& \times \frac{(1-\e/2)}{(1-\e(k+2)/4)(1-\e (k+2)/2)} \nonumber \\&&=  -\frac{k}{4(k+1)} \frac{ p^{2-\e(k+1)}}{\e^k} [1+(5k/4 +1)\e +O(\e^2)] \ea
The calculations from thereon are straightforward, we just have to expand in powers of $\e$ the various explicit functions which appear in (\ref{I},\ref{B}) and chose $Z$ to cancel all the poles in $\e$. For instance at order $g^3$ if we take
\be \label{Z} Z= 1- \frac{g^2}{N 144 \e} -\frac{g^3}{N 6^4 \e^2} (1-\e/4)  + O(g^4/N,1/N^2)\ee
we verify that $\G^{(2)}$ is finite up to this order, as renormalization theory implies,
\be \G^{(2)}(p) = p^2 - \frac{g^2}{144 N} p^2(2 \ln p -7/8) - \frac{9 g^3}{ 6^4N}p^2 \ln^2 p + O(g^4/N) . \ee
From (\ref{Z}) we obtain immediately  at leading order $1/N$
\be\label {ordre} \gamma (g) = \frac{1}{72 N} [ g^2 - \frac{1}{24} g^3 +O(g^4) ] \ee
which agrees for $N$ large with the result in \cite{ZJ}
\be \gamma(g) =\frac{N+2}{72 N^2 } g^2 [ 1 -\frac{N+8}{24N}g + \frac{5(-N^2+18 N +100)}{576 N^2} g^2] + \cdots \ee

We can proceed in this fashion to all orders in $g$, but it is tedious.  One can also sum the series  (\ref{S})  but it is not simple either to extract $Z$ from the sum.  
\be \G^{(2)} = Z p^2 -\frac {g^2 Z_1}{18 N} \int \frac{d^dq}{(p-q)^2}\frac{1}{\frac{1}{Z_1 B(q)} +g/6} \ee

Fortunately previous results on the $1/N$ expansion of this model allow us to  recover $\gamma(g)$,  at order $1/N$, to all orders in $g$ as explained in the next section. 

\section {Where are the renormalons?}

If instead of dimensional regularization we had stayed in four dimensions with an ultra-violet cut-off $\Lambda$ we could have computed the same diagrams The  bubble-diagram (\ref{bubble}),
\be \int_{\Lambda} \frac {d^4 v} {(v-q)^2 v^2}  \ee behaves as $ \ln {q/\Lambda}$ at small momentum. Inserted in the two-point  the  $k$ -th iterated bubble  behaves as $ \ln^k {q/\Lambda}$
and inserted in the two-point function it  yields the integral
\be  \int_{\Lambda} d^4 q [\frac{1}{(p-q)^2} - \frac{1}{q^2} ]\ln^k {q/\Lambda} \ee
where we have explicited the zero-momentum subtraction of the massless theory (which  automatically vanished in the minimal scheme).  The resulting integral  is porportional to $p^2$ and it yields an integral over $q$ which is infra-red singular in the $p$ small region. Taking $\ln q/\Lambda =-x$ the singular part is given by a power of a logarithmic singularity in $p$ , with a coefficient which behaves for large $x$ as $\int dx e^{-x} (-x)^k$, i.e. a factorial growth with alternating signs.  This is the argument for a perturbation expansion exhibiting (infra-red) renormalons \cite{MM}.

Our goal is to compute the renormalized correlation functions, the scaling limit of the theory in which distance are much larger than the lattice spacing $\Lambda^{-1}$.  We have seen in the previous section how complex is the interplay between the diagrams and the counter-terms and this  is not transparent in the above cut-off regularized theory.  So let us return to the results of the previous section within dimensional regularization. Looking back at the $k$-th order, i.e. the integral (\ref{Integral}), we see explicitly that the $k$ large and $\e$ small limits do not commute. The renormalization procedure is strictly defined as $\e$ goes to zero first, cancelling the poles in $\e$ through the poles coming from $Z$ and $Z_1$.  It is only after we have removed those singular terms that we may examine the asymptotic behavior for large $k$.   In the minimal subtraction scheme that we have followed here, this is done by cancelling all  the poles and multiple poles in $\e$ occuring in the results such as (\ref{Integral}) with the poles in $Z$ and $Z_1$.  

Could that produce the same renormalon-$k!$ ?    Indeed if we  return to  (\ref{Integral}) the factor $ \frac{1}{\e^k}  p^{2-\e(k+1)}$  will end up  expanded $k$ times in powers of $\e$ by the time all the subtractions manage to produce a finite four-dimensional theory (i.e, $\e =0$). This might yield  a behaviour of the two-point function at order $k$  with a term $p^2[-(k+1) \ln p]^k$  and of course $(-k)^k\simeq (-a)^k k!$, but this is far from obvious, the algebra could produce a $1/k!$  which would kill this would-be renormalon.  However our goal here is to understand the large orders of the $\e$-expansion and that relies on the expansion of the renormalization group $\gamma$ function. There we will see that there is no room for a renormalon large order behavior. 

\section {An explicit solution through earlier results}

Since the removal of the poles in $\e$  is increasingly more cumbersome when the order increases, fortunately we can call on  previous results on the large $N$ limit to bypass this long algebra.  In fact there are better ways of dealing with the $1/N$ expansion, like adding to the action a Lagrange multiplier $ \lambda ({\varphi}^2-\psi )$, replacing the quartic term in $\V$ by $\psi^2$ and tracing out the Gaussian $\V$'s. The expansion around the saddle-point of the resulting  $(\lambda, \psi)$  action yields the $1/N$-expansion \cite {ZJ,MM}. The reason for not  following  this procedure here is that we needed to stick to the minimal subtraction scheme. Several terms of the $1/N$ expansion of the  critical exponents have been computed for arbitrary dimensions, much more than what we needed here. In particular we find in \cite{MM} the critical exponent $\eta$ at leading order, and copy here the result
\be \label {eta}  \eta = \frac {1}{N} [ \frac{\e^2}{2}\frac{\G(1-\e)}{\G^3(1-\e/2)\G(1+\e/2) }\frac {1-\e}{ (1-\e/4)(1-\e/2)}]  + O(\frac {1}{N^2})\ee
But we know that $\eta = \gamma(g^{*})$ and $ g^{*} =6\e +O(1/N)$.  Therefore we obtain the RG function $\gamma (g)$ by replacing $\e$ by $g/6$ in $\eta$  providing the result to all orders in $g$
\be \label {ga} \gamma(g) = \frac {g^2}{72N} [ \frac{\G(1-g/6)}{\G^3(1-g/12)\G(1+g/12) }\frac {1-g/6}{ (1-g/24)(1-g/12)}]  + O(\frac {1}{N^2})\ee
This is the result valid to all orders  in $g$, first order in $1/N$, that we were looking for.  One verifies easily that this, expanded in $g$, reproduces  what we had  found  before at low order  (\ref{ordre}).

\section {Concluding remarks} 
\begin{itemize}
\item The result (\ref {ga}) is analytic in $g$ in the neighbourhood of the origin : it is meromorphic in $g$ with the closest singularity at $g=12$. Had the large orders of the expansion in $g$ be  growing factorially that the result would not be  analytic at $g=0$.  This is reminiscent of what is familiar in matrix models : the matrix integral is not analytic in $g$, the coefficient in $g\Tr M^4$. However if one considers the large $N$ limit, and the  successive terms of the $1/N$-expansion, every term of the expansion is analytic at $g=0$. It has been argued by previous authors that renormalons are not present in $\V^4_4$ \cite{S} : the explicit calculation performed here confirms  this position. 
\item The potentially dangerous renormalons do not show up in the final result
\item We have  not shown that other correlation functions, other than the one we have computed  here, could not show renormalons, but it seems likely, in view of what we did, that they are simply absent at first order in $1/N$ and we are inclined to believe that this remains true to all orders in a $1/N$ expansion.
\item We have not shown that renormalons would not show up at fixed $N$, but the argument in their favour being  a priori operative, but finally absent,  for the case that we have considered here, we see no reason to believe that they spoil the old result \cite {BLZ} on the large orders of the $\e$-expansion. 
\item The model that we have considered, was considered as a  candidate for infra-red renormalons. We have nothing to say on potential  UV renormalons as in gauge theories. 
\item Many contemporary scientists probably consider that the problem discussed here is obsolete ; who needs the $\e$-expansion given the magnificent precision of the conformal bootstrap \cite {CB}, which defeats earlier methods in their accuracy at predicting critical exponents ? However I believe that there are many problems of interest in the scaling region  for which the tools of conformal bootstrap are not (not yet?) available. For instance, the  universal scaling equation of state, still relies on expansions : it was my first  article  (with Wallace and Wilson) \cite {BWW} using the $\e$-expansion,  fifty years ago!
\end{itemize}

   \vskip 3mm 
 { \bf Acknowledgement}
 I thank Giorgio Parisi for a discussion which led me to reconsider this ancient story. 
 \vskip 2mm

  \vskip 5mm

 \end{document}